\documentclass[superscriptaddress, reprint, amsmath, amssymb, aps, prl, floatfix]{revtex4-2}

\usepackage{bm}

\usepackage{mathtools}
\usepackage{amsfonts}
\usepackage{amssymb}
\usepackage{dsfont}
\usepackage{amsthm}
\usepackage{nicefrac}
\usepackage[english]{babel}
\usepackage{graphicx}
\usepackage{amsmath}
\usepackage{algorithm}
\usepackage{algorithmic}
\usepackage{lipsum}
\usepackage{wrapfig}
\usepackage{float}
\usepackage{enumitem} 
\usepackage{color} 
\usepackage{hyperref}
\usepackage{url}
\usepackage{appendix}
\usepackage{changes}
\definecolor{blue-violet}{rgb}{0.54, 0.17, 0.89}

\useunder{\uline}{\ul}{}

\hypersetup{colorlinks = true, pdftitle = FALQON}

\usepackage{qcircuit}

\usepackage[normalem]{ulem}

\begin{document}

\title{Feedback-based quantum optimization}

\author{Alicia B. Magann} 

\affiliation{Quantum Algorithms and Applications Collaboratory, Sandia National Laboratories, Livermore, California 94550, USA}
\affiliation{Quantum Algorithms and Applications Collaboratory, Sandia National Laboratories, Albuquerque, New Mexico 87185, USA}
\affiliation{Department of Chemical \& Biological Engineering, Princeton University, Princeton, New Jersey 08544, USA}

\author{Kenneth M. Rudinger}
\affiliation{Quantum Algorithms and Applications Collaboratory, Sandia National Laboratories, Albuquerque, New Mexico 87185, USA}

\author{Matthew D. Grace}
\affiliation{Quantum Algorithms and Applications Collaboratory, Sandia National Laboratories, Livermore, California 94550, USA}

\author{Mohan Sarovar}
\affiliation{Quantum Algorithms and Applications Collaboratory, Sandia National Laboratories, Livermore, California 94550, USA}

\date{\today}

\begin{abstract}
It is hoped that quantum computers will offer advantages over classical computers for combinatorial optimization. Here, we introduce a feedback-based strategy for quantum optimization, where the results of qubit measurements are used to constructively assign values to quantum circuit parameters. We show that this procedure results in an estimate of the combinatorial optimization problem solution that improves monotonically with the depth of the quantum circuit. Importantly, the measurement-based feedback enables approximate solutions to the combinatorial optimization problem without the need for any classical optimization effort, as would be required for the quantum approximate optimization algorithm (QAOA). We experimentally demonstrate this feedback-based protocol on a superconducting quantum processor for the graph-partitioning problem MaxCut, and present a series of numerical analyses that further investigate the protocol's performance.
\end{abstract}

\maketitle

\paragraph*{Introduction.---} 
Combinatorial optimization has broad and high-value applications in many sectors of industry and science, including for optimization of logistics and supply chain, and drug discovery \cite{Papadimitriou_Steiglitz_1998}. Solving general combinatorial optimization problems is NP hard and most practical strategies involve developing good quality approximate solutions. Recently, there has been much interest in approximate solution of combinatorial optimziation problems through mapping to quantum systems, whereby the problem is encoded into an Ising Hamiltonian $H_{\textrm{p}}$ \cite{Lucas_2014}, such that the solution of problem is encoded in the ground state of $H_{\textrm{p}}$. Then methods such as quantum annealing \cite{Hauke_2020}, or within the quantum circuit model, the quantum approximate optimization algorithm (QAOA) \cite{2014arXiv1411.4028F}, are used to approximately prepare the ground state of $H_{\textrm{p}}$. Although there is no rigorous proof of an advantage to using such quantum techniques over classical approximation algorithms, it is widely believed that at some scale of problem such an advantage should exist.

We introduce a new approach to solving combinatorial optimization problems using quantum computers that operates through the use of parameterized quantum circuits and feedback, that is conditioned on qubit measurements at every quantum circuit layer, in order to determine the circuit parameter values at subsequent layers. 
This Feedback-based ALgorithm for Quantum OptimizatioN (FALQON) makes a direct connection to quantum Lyapunov control (QLC), a control strategy that uses feedback to identify the controls to drive the dynamics of a quantum system in a desired manner \cite{PhysRevLett.69.2172, doi:10.1063/1.467132, SUGAWARA1995113, OHTSUKI1998627, A902103E, doi:10.1063/1.1559680, 1272601,Mirrahimi2005ReferenceTT, doi:https://doi.org/10.1002/9780470431917.ch2}. Our approach works within the framework of circuit-model quantum computing, but avoids a critical challenge facing the scaling of QAOA, which is the difficulty of optimizing a large number of variational parameters. In fact, it was recently shown that under certain assumptions, this classical optimization problem is itself NP-hard for QAOA \cite{bittel2021training}. Our feedback-based approach circumvents the need for optimization of variational parameters by using information from iterative measurements.

In the following, we show that FALQON produces a monotonically improving estimate of the combinatorial optimization problem solution, with respect to the depth of the circuit. We then consider the application of FALQON towards solving the MaxCut problem, and present the results of an experimental demonstration on quantum hardware. This is followed by a series of numerical analyses that explore the performance of FALQON for MaxCut on 3-regular graphs. Finally, we examine the required number of repeated circuit evaluations and compare this to the requirements of QAOA in this context. We conclude with a discussion of the tradeoffs between FALQON and QAOA, outline the additional content in our companion paper \cite{CompanionPaper}, and look to the future.

\paragraph*{Feedback-based algorithm for quantum optimization.---}

\begin{figure*}
\includegraphics[width=\textwidth]{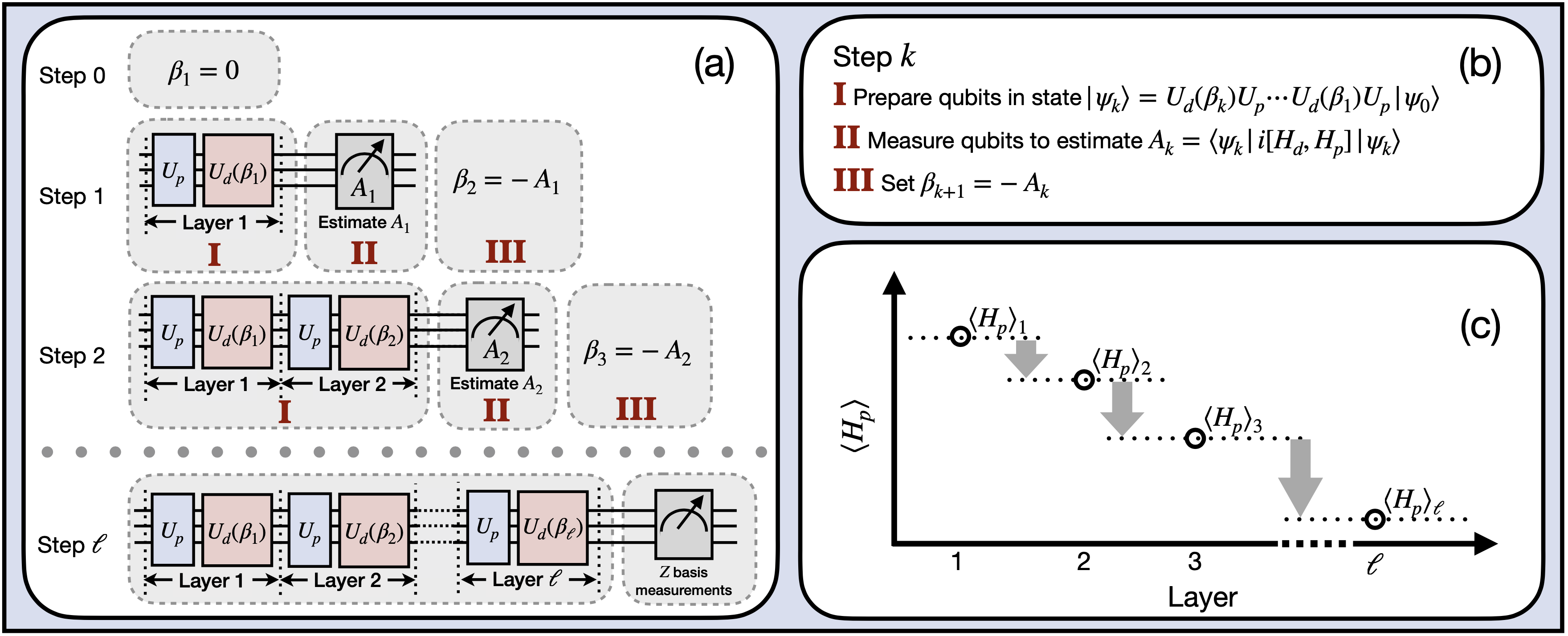}
\caption{\label{fig:concept}(a) The procedure for implementing FALQON. The initial step is to seed the procedure by setting $\beta_1=0$. The qubits are then initialized in the state $|\psi_0\rangle$, and a single FALQON layer is implemented to prepare $|\psi_1\rangle = U_{\textrm{d}}(\beta_1)U_{\textrm{p}}|\psi_0\rangle$. The qubits are then measured to estimate $A_1$, whose result is fed back to set $\beta_2 = -A_1$, up to sampling error. For subsequent steps $k = 2, \cdots, \ell$, the same procedure is repeated, as shown in (b): the qubits are initialized as $|\psi_0\rangle$, after which $k$ layers are applied to obtain $|\psi_k\rangle = U_{\textrm{d}}(\beta_{k}) U_{\textrm{p}}\cdots U_{\textrm{d}}(\beta_1) U_{\textrm{p}} |\psi_0\rangle$, and then the qubits are measured to estimate $A_{k}$, and the result is fed back to set the value of $\beta_{k+1}$. This procedure causes $\langle H_{\textrm{p}}\rangle$ to decrease layer-by-layer as per $\langle\psi_1| H_{\textrm{p}} |\psi_1\rangle\geq \langle\psi_2| H_{\textrm{p}} |\psi_2\rangle \geq \cdots \geq \langle\psi_\ell| H_{\textrm{p}} |\psi_\ell\rangle$, as shown in (c), such that the quality of the solution to the combinatorial optimization problem monotonically improves with circuit depth. The protocol can be terminated when the value of $\langle H_{\textrm{p}}\rangle$ converges or a threshold number of layers $\ell$ is reached. Then, after the final step, $Z$ basis measurements on $|\psi_\ell\rangle$ can be used to determine a best candidate solution to the combinatorial optimization problem of interest, by repeatedly sampling from the probability distribution over bit strings induced by $|\psi_\ell\rangle$ and selecting the outcome associated with the best solution.}
\end{figure*}
\twocolumngrid

We begin by considering a quantum system whose dynamics are governed by 
$    i\frac{d}{d t}|\psi(t)\rangle = (H_{\textrm{p}}+H_{\textrm{d}}\beta(t))|\psi(t)\rangle \, ,
$
where $|\psi(t)\rangle$ is the system state vector, we have set $\hbar = 1$, and $H_{\textrm{p}}$ and $H_{\textrm{d}}$ denote the (unitless) ``drift'' and ``control'' Hamiltonians, where the latter couples a scalar, time-dependent control function $\beta(t)$ to the system. We seek to minimize $\langle H_{\textrm{p}} \rangle = \langle \psi(t)| H_{\textrm{p}} |\psi(t)\rangle$ \footnote{In reference to QLC, $\langle H_{\textrm{p}}\rangle$ essentially serves as a control Lyapunov function, though in practice it may not be strictly positive or meet all of the requirements to be true Lyapunov function.}, and accomplish this by designing $\beta(t)$ such that 
\begin{equation}
\frac{d}{dt}\langle \psi(t)|H_{\textrm{p}}|\psi(t)\rangle(t)\leq 0, \quad \forall t \geq 0 \, .
\label{Eq:ddt}
\end{equation}

Evaluating the left-hand-side of Eq. (\ref{Eq:ddt}), we see that $\frac{d}{dt} \langle \psi(t)|H_{\textrm{p}}|\psi(t)\rangle = A(t) \beta(t)$, where $A(t)\equiv \langle\psi(t)| i[H_{\textrm{d}}, H_{\textrm{p}}] |\psi(t)\rangle$. There is significant flexibility in choosing $\beta(t)$ in order to satisfy Eq. (\ref{Eq:ddt}), i.e., we may take $\beta(t) = -w \, f(t,A(t))$,  for $w > 0$, where $f(t,A(t))$ is any continuous function with $f(t,0)=0$ and $A(t)f(t,A(t))>0$ for all $A(t)\neq 0$ \cite{lyapunovsurvey}. Here, we present results for $w = 1$ and $f(t,A(t)) = A(t)$, such that $\beta(t) = -A(t)$. In practice, we assign values to $\beta(t)$ as a feedback loop, where $\beta(t) = -A(t-\tau)$, and $\tau$ is a feedback loop time delay.

We now consider alternating, rather than concurrent, applications of $H_{\textrm{p}}$ and $H_{\textrm{d}}$, leading to a time evolution of the form $U = U_{\textrm{d}}(\beta_\ell) U_{\textrm{p}} \cdots  U_{\textrm{d}}(\beta_1) U_{\textrm{p}}$, where $U_{\textrm{p}}=e^{-i H_{\textrm{p}} \Delta t}$, $U_{\textrm{d}}(\beta_k)=e^{-i\beta_k H_{\textrm{d}}\Delta t}$, and $\beta_{k} = \beta(k\tau-\Delta t)$ for $k=1, 2, \cdots, \ell$ and $\tau = 2\Delta t$, such that after each period of $\Delta t$ the applied Hamiltonian alternates between $H_{\textrm{p}}$ and $H_{\textrm{d}}$. 
We note that for small $\Delta t$, this yields a Trotterized approximation to the continuous time evolution of the system. In this Trotterized framework, we again aim to satisfy Eq. (\ref{Eq:ddt}) by suitably choosing each value of $\beta_k$. We note that during the time intervals when $H_p$ is applied, $\frac{d}{dt}\langle  H_{\textrm{p}} \rangle(t)=0$; although its value doesn't change, the eigenstates of $H_p$ do accumulate phases during this time, which impact the ensuing dynamics. Meanwhile, during the time intervals when $H_d$ is applied, we recover the same result that $\frac{d}{dt} \langle H_{\textrm{p}}\rangle = A(t) \beta(t)$. Consequently, we can ensure that Eq. (\ref{Eq:ddt}) is satisfied by utilizing the same feedback law, given by $\beta_{k+1}= -A_k$, where $A_k = \langle\psi_k| i[H_{\textrm{d}}, H_{\textrm{p}}] |\psi_k\rangle$ \footnote{Estimating each $A_k$ can be accomplished by first expanding it in the Pauli operator basis as $A_k = \langle \psi_k|i[H_{\textrm{d}}, H_{\textrm{p}}]|\psi_k\rangle = \Sigma_{j=1}^N \alpha_j\langle\psi_k| P_j|\psi_k\rangle$, where $\alpha_j$ are scalar coefficients and $P_j$ are Pauli strings, and then measuring the expectations of each $P_j$ in order to evaluate the weighted sum. The value of $N$ depends on $H_{\textrm{p}}$ and $H_{\textrm{d}}$. For the MaxCut examples we consider, $N\leq n(n-1)$}. In this setting, it is always possible to select $\Delta t$ small enough such that Eq. (\ref{Eq:ddt}) is satisfied \cite{CompanionPaper}. However, if $\Delta t$ is chosen to be too large, Eq. (\ref{Eq:ddt}) will be violated. Based on this framework, the FALQON algorithm is presented in Fig.~\ref{fig:concept}. The key feature of FALQON is that it is a constructive, optimization-free procedure for assigning values to each $\beta_k$ according to a feedback law. And by design, the enforcement of Eq. (\ref{Eq:ddt}) ensures that the quality of the solution to the combinatorial optimization problem under consideration (quantified by $\langle H_p \rangle$) improves monotonically with respect to the depth of the circuit, $k$. 

The circuits used in QAOA have the same alternating structure as those in FALQON, albeit with additional parameters $\gamma_1, \cdots, \gamma_{\ell}$ that enter into $U_{\textrm{p}}$, such that $U_{QAOA} = U_{\textrm{d}}(\beta_\ell) U_{\textrm{p}}(\gamma_\ell) \cdots U_{\textrm{d}}(\beta_1) U_{\textrm{p}}(\gamma_1)$. Then, the solution to the original combinatorial optimization problem is sought by minimizing $\langle \psi (\vec{\gamma}, \vec{\beta})|H_{\textrm{p}}|\psi(\vec{\gamma}, \vec{\beta})\rangle$ over the set of $2\ell$ circuit parameters $\vec{\gamma} = (\gamma_1, \cdots, \gamma_\ell)$ and $\vec{\beta} = (\beta_1, \cdots, \beta_\ell)$ using a classical processor, where $|\psi(\vec{\gamma}, \vec{\beta})\rangle = U_{QAOA} |\psi_0\rangle$. However, we emphasize that FALQON is conceptually distinct from QAOA. Namely, QAOA seeks to minimize $\langle H_{\textrm{p}} \rangle$ by classically optimizing over all parameters $\vec{\gamma},\vec{\beta}$ simultaneously, while FALQON seeks to minimize $\langle H_{\textrm{p}} \rangle$ over a sequence of quantum circuit layers, guided by qubit measurement-based feedback, \emph{without classical optimization}. 

\paragraph*{Applications to MaxCut.---}

We now consider the application of FALQON towards a quintessential combinatorial optimization problem: MaxCut, which aims to identify a graph partition that maximizes the number of edges in a graph that are cut. For an unweighted graph $\mathcal{G}$, with $n$ nodes and edge set $\mathcal{E}$, the MaxCut problem Hamiltonian is defined on $n$ qubits as
$H_{\textrm{p}} = -\sum_{i,j \in \mathcal{E}} \frac{1}{2} \big(1-Z_i Z_j\big)\,,$
while $H_{\textrm{d}}$ has the standard form $H_{\textrm{d}}=\sum_{j=1}^nX_j$, such that  $i[H_{\textrm{d}},H_{\textrm{p}}] = \sum_{i,j \in \mathcal{E}} Y_i Z_j + Z_i Y_j$, where $X_j$, $Y_j$, and $Z_j$ denote the Pauli operators acting on qubit $j$. As such, evaluating the feedback law $\beta_{k+1}=-A_k = -\langle\psi_k|i[H_{\textrm{d}},H_{\textrm{p}}]|\psi_k\rangle$ in this setting involves measurements of maximally $n(n-1)$ two-qubit Pauli strings.

\begin{figure}[t]
\includegraphics[width=1.0\columnwidth]{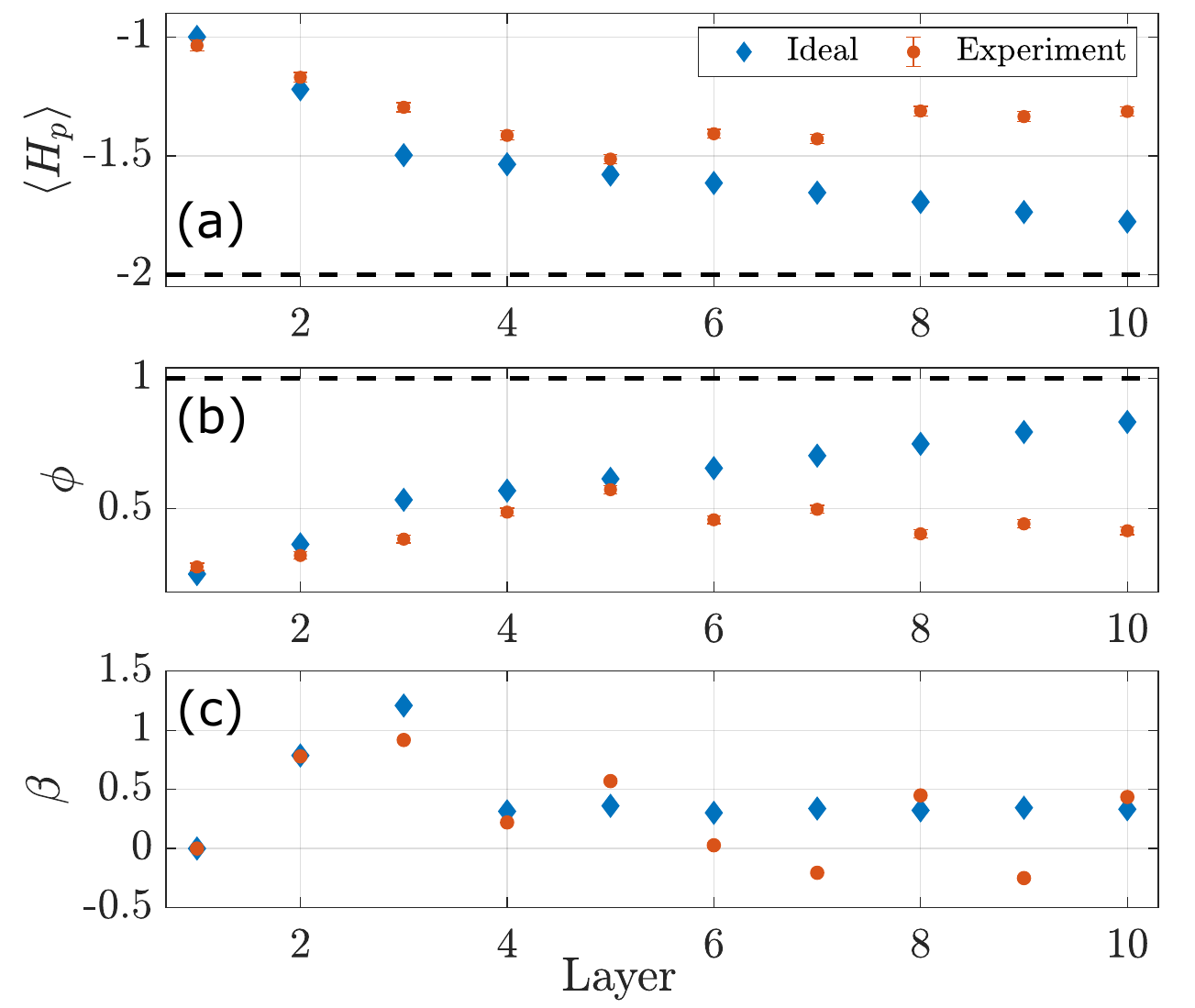}
\caption{{Results from experimental implementation of $\ell=10$ layers of FALQON on a superconducting quantum processor. For this demonstration, FALQON is applied to an $n=3$ qubit instance of MaxCut on an unweighted graph. Panel (a) shows that FALQON is successful in achieving a monotonic decrease of $\langle H_p\rangle$ over layers $k=1,\cdots,5$ in this experiment (orange point markers), noting that the global minimum value for this problem instance is $\langle H_p\rangle_{\text{min}}=-2$ (dashed black line). In addition, in panel (b) a monotonic increase of the probability, $\phi$, of measuring the two degenerate ground states is also observed up to layer $k=5$ (orange point markers). The error bars in (a) and (b) indicate the standard error. The values of $\beta$ are plotted in (c). In (a)-(c), the blue point markers correspond to ideal results computed numerically.  } }
\label{Hardware}
\end{figure}

As a proof-of-principle, in Fig. \ref{Hardware} we present the results of an experimental demonstration of FALQON on a superconducting quantum processor for a simple instance of the MaxCut problem. In particular, we considered an instance of MaxCut on an unweighted graph composed of $n=3$ nodes connected by two edges, such that $H_p = -\tfrac{1}{2}(2-Z_1Z_2-Z_2Z_3)$ and $i[H_{\textrm{d}},H_{\textrm{p}}] = Y_1 Z_2 + Z_2 Y_1+Y_2 Z_3 + Z_3 Y_2$. The experiment was performed on the publicly accessible \texttt{ibmq$\_$manila} processor and utilized three qubits with nearest-neighbor connectivity matching that of the graph under consideration. In this setting, $\ell=10$ steps of FALQON were performed according to the procedure outlined in Fig. \ref{fig:concept}, selecting $\Delta t = 0.2$. At each step, one circuit was implemented in order to estimate $\langle H_p\rangle_k$ natively in the computational basis. Two additional circuits were implemented in order to estimate the terms in $A_k$. For each circuit, the qubits were initialized in the ground state of $H_{\textrm{d}}$, and $m=1024$ shots were taken.

As shown in Fig. \ref{Hardware}(a), FALQON was successful in achieving a monotonic decrease of $\langle H_p\rangle$ in this experiment up to layer five (orange point markers). FALQON also achieves a monotonic increase in the success probability of measuring the two degenerate ground states, denoted by $\phi$, as shown in Fig. \ref{Hardware}(b). The error bars in Fig. \ref{Hardware}(a) and (b) present the standard error of the mean, which estimates how much the reported $\langle H_p\rangle_k$ and $\phi_k$ may deviate from their true values due to finite sampling. Finally, the associated values of $\beta$, determined according to the feedback law $\beta_{k+1} = -A_k$, are plotted in Fig. \ref{Hardware}(b).

Past layer 5, it is evident that FALQON is no longer able to decrement $\langle H_p\rangle$ using this hardware platform, despite exhibiting a continued monotonic decrease in associated noise-free numerical simulations (blue point markers). This reveals the limitations that hardware noise presents for this problem instance. Looking ahead, we are optimistic that continuous improvements to quantum hardware will pave the way towards applications of FALQON to increasingly complex combinatorial optimization problems.

\begin{figure*}
\centering
\includegraphics[width=2.0\columnwidth]{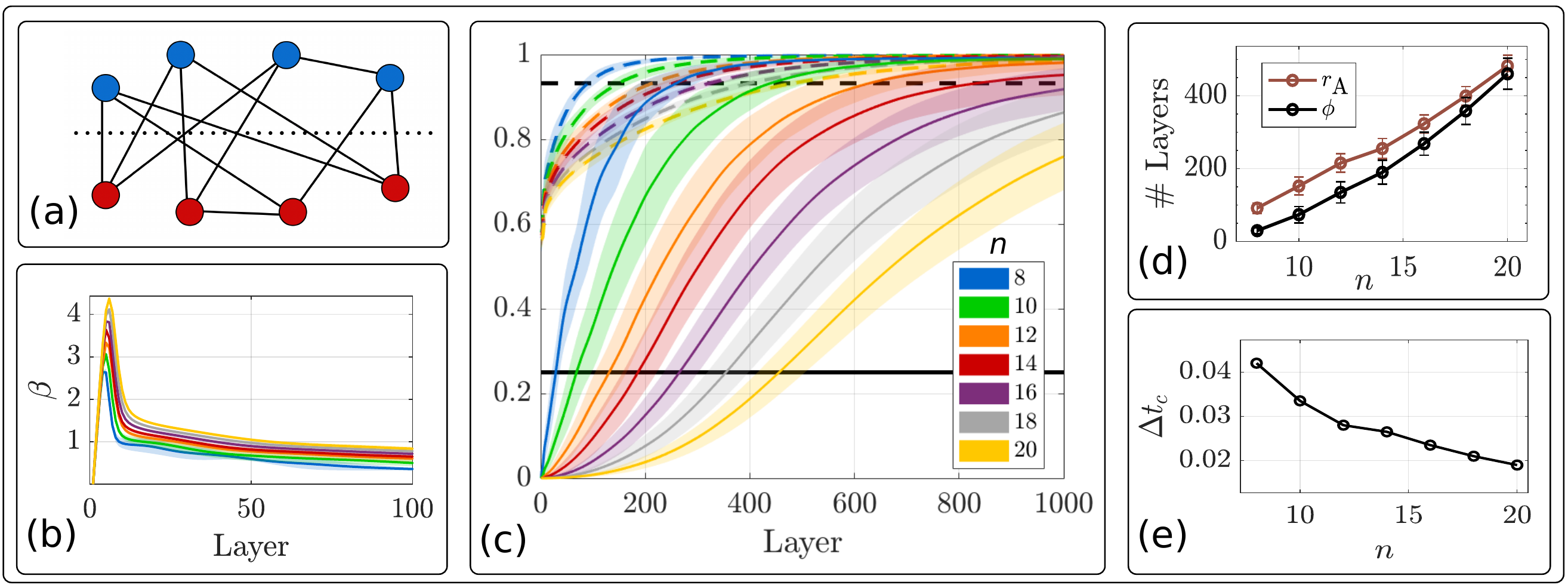}
\caption{(a) Pictorial representation of MaxCut on a 3-regular graph with 8 vertices. (b) Mean $\beta$ values are plotted as a function of layer for different $n$ values, with shading showing the standard deviations. (c) The performance of FALQON, as quantified by the approximation ratio (dashed curves) and the success probability of measuring the degenerate ground state (solid curves) is shown for different values of $n$. (d) The mean number of layers needed to achieve the reference values of $r_{\textrm{A}} = 0.932$ (dashed curve) and $\phi = 0.25$ (solid curve) is shown; error bars report the associated standard deviation. (e) The critical $\Delta t$ values for different problem sizes are plotted.}
\label{3reg}
\end{figure*}

In the interim, we explore how FALQON performs on larger instances of MaxCut through a series of noise-free numerical illustrations. These illustrations consider unweighted, connected 3-regular graphs with $n \in \{8, 10, \cdots, 20\}$ vertices. For $n \in \{8, 10\}$ we consider all nonisomorphic graphs; for $n \in \{12, 14, \cdots, 20\}$ we consider 50 randomly-generated, nonisomorphic graphs. In our simulations, the qubits are initialized in the ground state of $H_{\textrm{d}}$, and the performance of FALQON is quantified using the mean and standard deviations (over the problem instances) of two figures of merit: the approximation ratio, $r_{\textrm{A}} = \langle H_{\textrm{p}} \rangle / \langle H_{\textrm{p}} \rangle_{\min}$ and the success probability of measuring the (potentially degenerate) ground state(s) $\{|q_{0,i}\rangle\}$, $\phi=\sum_i|\langle \psi| q_{0,i} \rangle|^2$. We relate the performance to two reference values: $r_{\textrm{A}} = 0.932$, corresponding to the highest approximation ratio that can currently be guaranteed using a classical approximation algorithm (i.e., the algorithm of Goemans and Williamson \cite{goemans1995improved}), and $\phi = 0.25$, which implies that on average, four repetitions will be needed in order to obtain a sample corresponding to the ground state.  Our only free parameter is the time step $\Delta t$, which is tuned to be as large as possible, a value we call the \emph{critical} $\Delta t$ and denote by $\Delta t_c$, as long as the condition in Eq.~(\ref{Eq:ddt}) is met for all problem instances considered. Our results are collected in Fig.~\ref{3reg}. In Fig.~\ref{3reg}(b), the mean values of $\beta_1, \beta_2, \cdots$ are plotted as a function of layer for different values of $n$, according to the legend in Fig.~\ref{3reg}(c), with the shading representing the standard deviation. We find that with increasing $n$, the shape of the resultant $\beta$ curves follows a clear trend, and the standard deviation decreases. In Fig.~\ref{3reg}(c), the associated $r_{\textrm{A}}$ and $\phi$ results are shown (dashed and solid curves, respectively), and the associated reference values are plotted in black. For the cases considered here, we find that FALQON consistently leads to monotonic convergence towards very high $r_{\textrm{A}}$ and $\phi$ values as a function of layer. To determine how the requisite circuit depths scale with the problem size, in Fig.~\ref{3reg}(d) we plot the average number of layers required to achieve the reference values of $r_{\textrm{A}}$ and $\phi$ as a function of $n$. Finally, in Fig.~\ref{3reg}(e) we plot $\Delta t_c$ for each value of $n$ under consideration. The scaling of the required number of layers and $\Delta t_c$ seems nearly linear, even up to $n = 20$, indicating a favorable runtime scaling of the FALQON algorithm, at least for this class of MaxCut problems. We remark that in addition to the analyses presented here, we also tested the performance of FALQON on weighted 3-regular graphs, and identified instances where the $r_{\textrm{A}}$ and $\phi$ convergence is enhanced by introducing one of three possible heuristic modifications to the FALQON algorithm. Details can be found in Appendix A.

In our companion paper \cite{CompanionPaper}, we present a sampling complexity comparison between FALQON and QAOA in the context of MaxCut, as quantified by the total number of samples (i.e., circuit repetitions) that are required, denoted $N_{s}$. When a gradient algorithm is used for QAOA, $N_{s}^{QAOA}  = \mathcal{O}(mq(\ell)\ell)$, where $m$ denotes the number of samples needed to estimate the expectation value of a two-qubit Pauli string $P_j$, and for simplicity, $m$ is assumed to be independent of $P_j$ and $q$ denotes the number of classical optimization iterations. For gradient-free methods, $N_{s}^{QAOA} =\mathcal{O}(mq(\ell))$. Meanwhile, in FALQON we find $N_{s}^{FALQON} = \mathcal{O}(md\ell)$, where $d$ denotes the degree of the graph. This suggests that FALQON has a more favorable sampling complexity than QAOA for cases where the number of QAOA optimization iterations $q(\ell)$ exceeds $d\ell$ in general, or $ d$ when a gradient algorithm is utilized. Further details can be found in \cite{CompanionPaper}.

\paragraph*{Discussion and outlook.--}

We have introduced FALQON as a constructive, feedback-based algorithm for solving combinatorial optimization problems using quantum computers. Importantly, FALQON performs optimization without the need for an expensive classical optimization loop. We have demonstrated its performance on current quantum hardware and provided numerical analyses of its performance towards finding the maximum cut of regular graphs. By studying the performance with respect to layer and the problem size $n$, our numerical analyses show that FALQON converges to very high approximation ratios and success probabilities with a favorable scaling of resources with respect to $n$, suggesting that FALQON may be a useful heuristic algorithm for this class of problems.

Our findings also suggest that FALQON can require relatively deep circuits in order to achieve this convergence, relative to the shallow circuits typically considered in QAOA. In our companion article \cite{CompanionPaper}, we provide an in-depth analysis of the tradeoffs in the performance and resource requirements of FALQON and QAOA, and discuss the resource regimes where each of these methods can be expected to offer advantages. In short, we expect QAOA to be favorable in settings where suitable classical optimization resources are available and quantum resources are restricted to the regime of shallow circuits. Meanwhile, FALQON performs well for deep circuits and does not require any classical optimization resources, meaning that there is no rising classical cost as the quantum circuit depth is increased. This indicates that in settings where deep circuits are feasible, FALQON is a new heuristic that could offer a considerable advantage.

In addition, our companion paper \cite{CompanionPaper} also includes the following other important elements. (1) We present an analysis of convergence criteria for the algorithm. (2) For the analysis presented here we have assumed ideal, noiseless access to the expectation values $A_k$ that dictate the feedback signal, $\beta_k$, however, in \cite{CompanionPaper} we show that FALQON is robust to noise in this quantity stemming from finite sample estimates of these expectations. This robustness ultimately stems from the flexibility in choosing $\beta$ to satisfy Eq. (\ref{Eq:ddt}). (3) We compare the performance of FALQON and QAOA for a fixed number of circuit repetitions, and we also explore how FALQON can also be used to seed QAOA by identifying a set of initial QAOA parameters that can serve as the starting point for subsequent iterative optimization. We show that this seeding procedure is useful in settings with limited circuit depth, in cases where FALQON fails to converge on its own, and in cases where QAOA fails to converge on its own due to difficulty with effective initialization of the optimization procedure. (4) We numerically demonstrate FALQON on weighted MaxCut, detail some possible extensions to the protocol, and analyze the relationship between FALQON and quantum annealing protocols.

Finally, we note that FALQON can be applied to combinatorial optimization problems beyond MaxCut, \emph{e.g.,} \cite{PennyLane}, and could have broader implications for quantum variational algorithms. That is, it is possible to develop feedback-based alternatives of variational ansatz\"e for other applications such as electronic structure or machine learning \cite{Cerezo_2020}, and these would have the benefit of needing no classical optimization resources, at the cost of requiring measurements whose results condition the feedback.

\begin{acknowledgments}
\paragraph*{Acknowledgments.--}
We gratefully acknowledge discussions with C. Arenz, L. Brady, L. Cincio, T.S. Ho, L. Kocia, O. Parekh, H. Rabitz, and K. Young. MS also thanks M. Fenech. This work was supported by the U.S. Department of Energy, Office of Science, Office of Advanced Scientific Computing Research, under the Quantum Computing Application Teams program. A.B.M. also acknowledges support from the U.S. Department of Energy, Office of Science, Office of Advanced Scientific Computing Research, Department of Energy Computational Science Graduate Fellowship under Award Number DE-FG02-97ER25308, as well as support from Sandia National Laboratories’ Laboratory Directed Research and Development Program under the Truman Fellowship. M.D.G. also acknowledges support from the U.S. Department of Energy, Office of Science, Advanced Scientific Computing Research, under the Accelerated Research in Quantum Computing (ARQC) program. SAND2022-14511 J.

This article has been authored by an employee of National Technology \& Engineering Solutions of Sandia, LLC under Contract No. DE-NA0003525 with the U.S. Department of Energy (DOE). The employee owns all right, title and interest in and to the article and is solely responsible for its contents. The United States Government retains and the publisher, by accepting the article for publication, acknowledges that the United States Government retains a non-exclusive, paid-up, irrevocable, world-wide license to publish or reproduce the published form of this article or allow others to do so, for United States Government purposes. The DOE will provide public access to these results of federally sponsored research in accordance with the DOE Public Access Plan {https://www.energy.gov/downloads/doe-public-access-plan}. This paper describes objective technical results and analysis. Any subjective views or opinions that might be expressed in the paper do not necessarily represent the views of the U.S. Department of Energy or the United States Government.

This report was prepared as an account of work sponsored by an agency of the United States Government. Neither the United States Government nor any agency thereof, nor any of their employees, makes any warranty, express or implied, or assumes any legal liability or responsibility for the accuracy, completeness, or usefulness of any information, apparatus, product, or process disclosed, or represents that its use would not infringe privately owned rights. Reference herein to any specific commercial product, process, or service by trade name, trademark, manufacturer, or otherwise does not necessarily constitute or imply its endorsement, recommendation, or favoring by the United States Government or any agency thereof. The views and opinions of authors expressed herein do not necessarily state or reflect those of the United States Government or any agency thereof.

\end{acknowledgments}

\bibliography{bib}

\section{Appendix A: Heuristic improvements}

Our numerical illustrations involving MaxCut on unweighted 3-regular graphs show that FALQON converges to very high approximation ratios and success probabilities. However, we also tested the performance of FALQON on weighted graphs, and were able to identify problem instances where $r_{\textrm{A}}$ appears to converge to very high values, while the convergence of $\phi$ is less favorable, i.e., $\beta \rightarrow 0$ prior to $\phi \rightarrow 1$, indicating that $\beta$ tends to zero prematurely. Like behavior has been found in numerical studies of QAOA, where the inclusion of edge weights leads to the appearance of many poor-quality local minima in the optimization landscape \cite{shaydulin2022parameter}. To cope with these situations, we introduce three heuristic modifications that can be used to enhance the performance of FALQON.

The first modification is to incorporate random ``kicks'' into $\beta$.  For some $\beta_c$ of our choosing, for all $\beta_k < \beta_c$, with probability $P_k$ we set $\beta_k = \beta_c$. We choose $\beta_c = 1$ and $P_k = (1-\beta_k) \alpha_k$, where $\alpha_k = 0.1 \sin^2(\frac{\pi k}{2\ell} -\frac{\pi}{2})$ is designed to decrease to zero as a function of circuit depth.

We also consider a second heuristic inspired by QLC, where the use of a reference perturbation $\lambda(t)$ in the control $\beta(t)$, such that $H(t) = H_{\textrm{p}} + (\lambda(t) + \beta(t)) H_{\textrm{d}}\,,$ has been considered in order to improve convergence \cite{BEAUCHARD2007388, doi:10.1002/rnc.1748, lyapunovsurvey}. In this setting, we may define {System (a)} as a system with drift Hamiltonian $H_{\textrm{p}}$ and control Hamiltonian $H_{\textrm{d}}$, and {System (b)} as the perturbed system with drift Hamiltonian $H_{{\textrm{p}},(b)}(t) \equiv H_{\textrm{p}} + \lambda(t)H_{\textrm{d}}$ and control Hamiltonian $H_{\textrm{d}}$. Then, the time-derivative of $\langle\psi(t)| H_{{\textrm{p}},2}(t) |\psi(t)\rangle$, allows us to define $\beta(t) = -A(t)$ as usual to ensure $\frac{d}{dt}\langle\psi(t)| H_{{\textrm{p}},2}(t) |\psi(t)\rangle \leq 0$. Within this framework, if {System (b)} converges asymptotically to the ground state of $H_{{\textrm{p}},2}(t)$, and if $\lambda(t)=0$ when this occurs, then {System (b)} becomes {System (a)}, such that the ground state of $H_{{\textrm{p}},(b)}(t)$ is also the ground state of $H_{\textrm{p}}$, and the method has converged successfully to the desired state. In practice, $\lambda(t)$ can be chosen to be a slowly-varying reference function that tends to 0 as $t\rightarrow \infty$. This framework can be translated into a modified version of FALQON by discretizing as before; for our numerical illustrations, we chose $\lambda_k = \alpha_k$.

\begin{figure}[b]
\includegraphics[width=1.0\columnwidth]{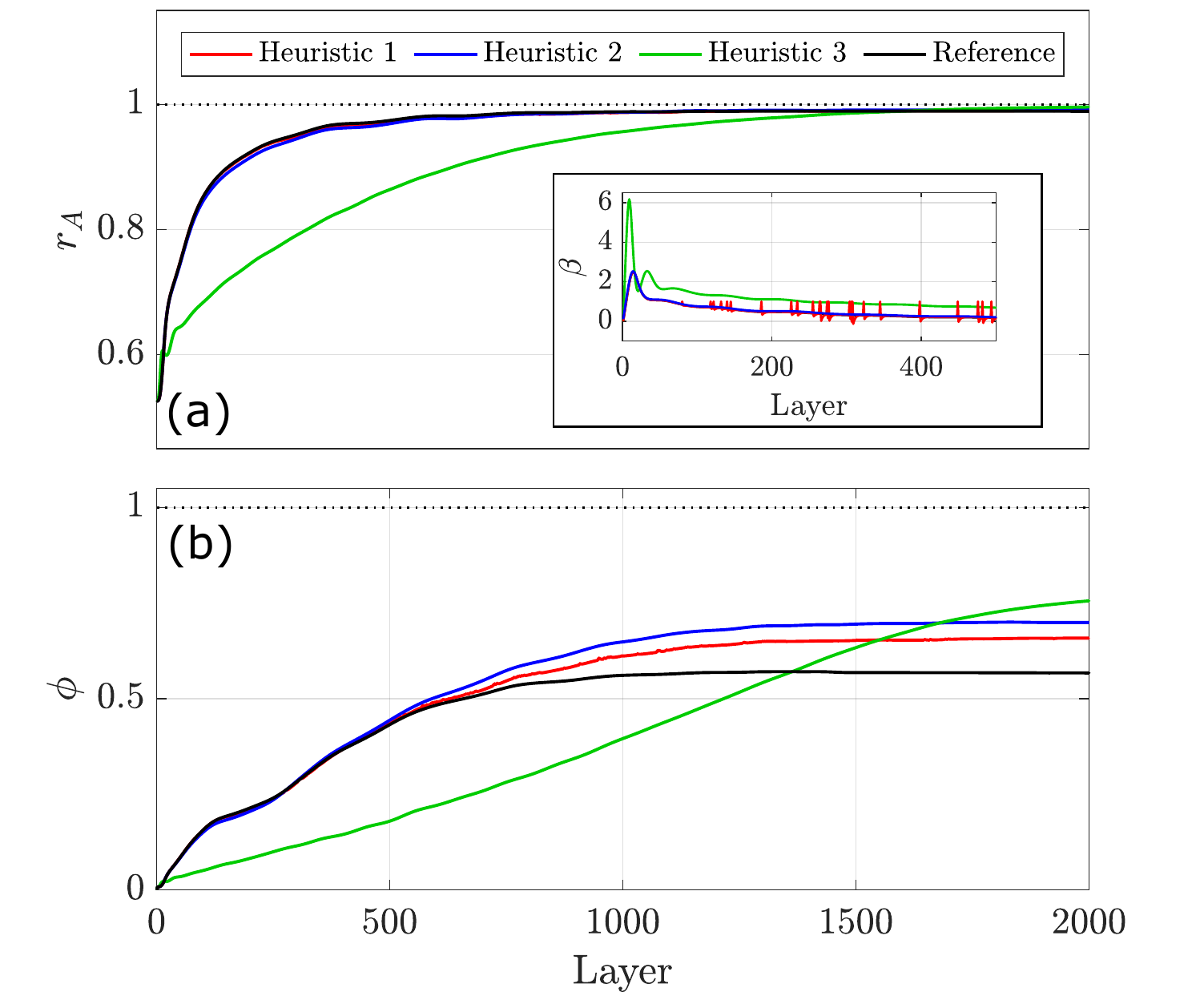}
\caption{Standard FALQON (black) is compared against the three heuristic modifications, which incorporate random kicks (red), a reference perturbation (blue), and three iterations of the iterative QLC procedure (green), for MaxCut on a weighted, 4-regular graph with $n=8$ vertices. The $r_{\textrm{A}}$ and $\phi$ results are shown in (a) and (b), respectively; associated values of $\beta$ are plotted for the first 400 layers in the inset.}
\label{Heuristics}
\end{figure}

Then, using this second heuristic as a baseline, we can define a third heuristic that uses an iterative QLC procedure to successively refine $\beta$ in a manner that is free of any classical optimization \cite{Mirrahimi2005ReferenceTT}. The procedure begins by implementing the standard FALQON framework and obtaining a set of $\beta$ values for $\ell$ layers. Then, these initial $\beta = \beta^{(0)}$ values are set as a reference perturbation $\lambda^{(1)}$, and a new set of $\beta^{(1)}$ values are obtained using the second heuristic approach described above. Then, a new reference perturbation is defined as $\lambda^{(2)} = \lambda^{(1)}+\beta^{(1)}$, and the process is repeated. If $\ell$ is selected to be large enough such that $\beta_\ell = 0$, this iterative procedure guarantees a monotonic improvement of $\langle H_p\rangle$ with respect to iteration. For further details, we refer the reader to our companion paper \cite{CompanionPaper}.

To illustrate these heuristic modifications, in Fig.~\ref{Heuristics} we present the performance of FALQON with and without these modifications when solving a MaxCut problem on a weighted, 4-regular graph with $n = 8$ nodes using $\Delta t = 0.08$, where the edge weights are drawn from a uniform distribution between 0 and 1. 

\end{document}